\documentclass{PoS}

\title{Heavy-Quark Masses from the Fermilab Method in Three-Flavor Lattice QCD}

\ShortTitle{Heavy-Quark Masses from the Fermilab Method}

\author{\speaker{Elizabeth D. Freeland}\\
         The School of the Art Institute of Chicago, Chicago, Illinois, USA\\
         Fermi National Accelerator Laboratory, Batavia, Illinois, USA$^\dag$\\
        E-mail: \email{eliz@fnal.gov}}
        
\author{Andreas~S.~Kronfeld, James~N.~Simone, and 
   Ruth~S.~Van~de~Water\\
        Fermi National Accelerator Laboratory, Batavia, Illinois, USA
         \thanks{Operated by Fermi Research Alliance, LLC, under Contract No.~DE-AC02-07CH11359 with the United States Department of Energy.}
        }

 \author{Fermilab Lattice Collaboration with the MILC Collaboration}

\abstract{
We report on heavy quark mass calculations using Fermilab heavy quarks.
Lattice calculations of heavy-strange meson masses are  combined with one-loop (automated) lattice perturbation theory to arrive at the quark mass.  Mesons are constructed from Fermilab heavy quarks and staggered light quarks.  We use the MILC ensembles at three lattice spacings and sea quark mass ratios of $m_{\rm u,d} / m_{\rm s} = 0.1$ to 0.4.  Preliminary results for the bottom quark are given in the potential subtracted scheme.}

\FullConference{The XXV International Symposium on Lattice Field Theory\\
		 July 30 - August 4 2007\\
		 Regensburg, Germany}

\begin{document}

% OVERVIEW
\section {Overview}

%Goal
An important contribution of lattice QCD to phenomenology is the calculation of quark masses.  Here, we discuss the mass calculation of the heavy quarks bottom and charm.  Our method combines Monte Carlo calculations of heavy-light mesons with lattice perturbation theory. 
This first section provides an overview of the calculation
which uses the Fermilab method for heavy quarks. 
Sections~\ref{nonpt} and~\ref{pt} review the non-perturbative and perturbative aspects, respectively.  We conclude with preliminary results for the bottom quark mass.

% pole mass(es)
Because the lattice violates Lorentz (Euclidean) invariance, the energy-momentum relationship, 
\begin{equation}
\label{dispreln}
 	E^2({\bf p}) = m_1^2 + \frac{m_1}{m_2} \, {\bf p}^2 + \ldots,
\end{equation}
is distorted~\cite{El-Khadra:1996mp}.
The  mass $m_1$ is called the rest mass and $m_2$  the kinetic mass.  They  are defined, for quarks or hadrons, as~\cite{El-Khadra:1996mp} 
\begin{equation}
\label{polemasses}
 	m_1  \equiv E({\bf 0})  \qquad \qquad
	m_2  \equiv \left( \frac{\partial^2 E}{\partial p_i^2} \right)_{\bf p=0}^{-1}.
\end{equation}
Both $m_1$ and $m_2$ provide a means for determining the quark mass.
The two methods must yield the same result in the continuum limit, providing a cross-check on our results.
%
%Two pole masses leads to two methods for calculating the quark mass which we refer to as the rest-mass and kinetic-mass methods.

%rest mass method
The first method uses the meson binding energy  to arrive at the quark mass.   For the Fermilab method, it has been shown that~\cite{Kronfeld:2000ck}, 
\begin{equation}
\label{bindingE}
  	M_1 - m_1 = M_{\rm expt} - m_{\rm pole}
  		\qquad  \rightarrow  \qquad  
	m_{\rm pole} =  m_1 + (M_{\rm expt} - M_1)
\end{equation}
where $M_1$ is the heavy-light meson rest mass calculated on the lattice, $M_{\rm expt}$ is the experimentally measured  meson mass, $m_1$ is the lattice heavy-quark pole rest mass and $m_{\rm pole}$ is the continuum quark pole mass.
This equation holds up to 
discretization errors in operators of dimension six and higher, and to truncation error in $m_1$, when that is defined perturbatively.
 Specifically, the leading mismatch in the Lagrangians 
 %between the continuum and lattice heavy-quark-effective-theory Lagrangians 
 comes from the hyperfine interaction, the Darwin term and the spin-orbit interaction~\cite{Kronfeld:2000ck}.  
Equation~(\ref{bindingE}) also holds for spin-averaged mesons,  $\overline{M}_1$ and $\overline{M}_{\rm expt}$, so spin-dependent discretization effects 
%from the chromomagnetic interaction are eliminated for spin-averaged meson masses, $\overline{M}_1$,  we use them here.  
can be eliminated by using them.
This leaves the Darwin term as the leading source of discretization errors.
Calculating the left-hand side of Eq.~(\ref{bindingE}) on the lattice and using PDG~\cite{PDG06} values for $\overline{M}_{\rm expt}$, we arrive at a value for $m_{\rm pole}$.  We refer to this method as the rest-mass method.

 %kin mass method
Alternatively, we can use the kinetic mass which the Fermilab method identifies with the physical quark mass, $a m_2 = a m_{\rm pole}$.
We use the ratio $\overline{M}_{\rm expt}  /  a \overline{M}_2$ to set the (inverse) lattice spacing.  
The quark mass is then
\begin{equation}
\label{kinmass}
 	m_{\rm pole} =  am_2 \; \frac {\overline{M}_{\rm expt} }{ a \overline{M}_2}  
		= \frac{am_2} { a \overline{M}_2}   \; \overline{M}_{\rm expt}.
\end{equation}
The advantage of setting the lattice spacing this way can be seen in the second equality.
It shows how a mistuning of the heavy quark mass cancels in the ratio $am_2  / a \overline{M}_2$.
%where the ratio $am_2  / a \overline{M}_2$ allows for cancellations of the effects of mistuning of the heavy quark.
%The second equality shows how this formulation cancels the effects of mistuning of the heavy quark.    
We refer to Eq.~(\ref{kinmass}) as the kinetic-mass method.

% NON_PERTURBATIVE ELEMENTS
\section {Non-perturbative Elements} \label{nonpt}
%
% FIGURE
\begin{figure}
	\begin{center}
      		\includegraphics[scale=0.26]{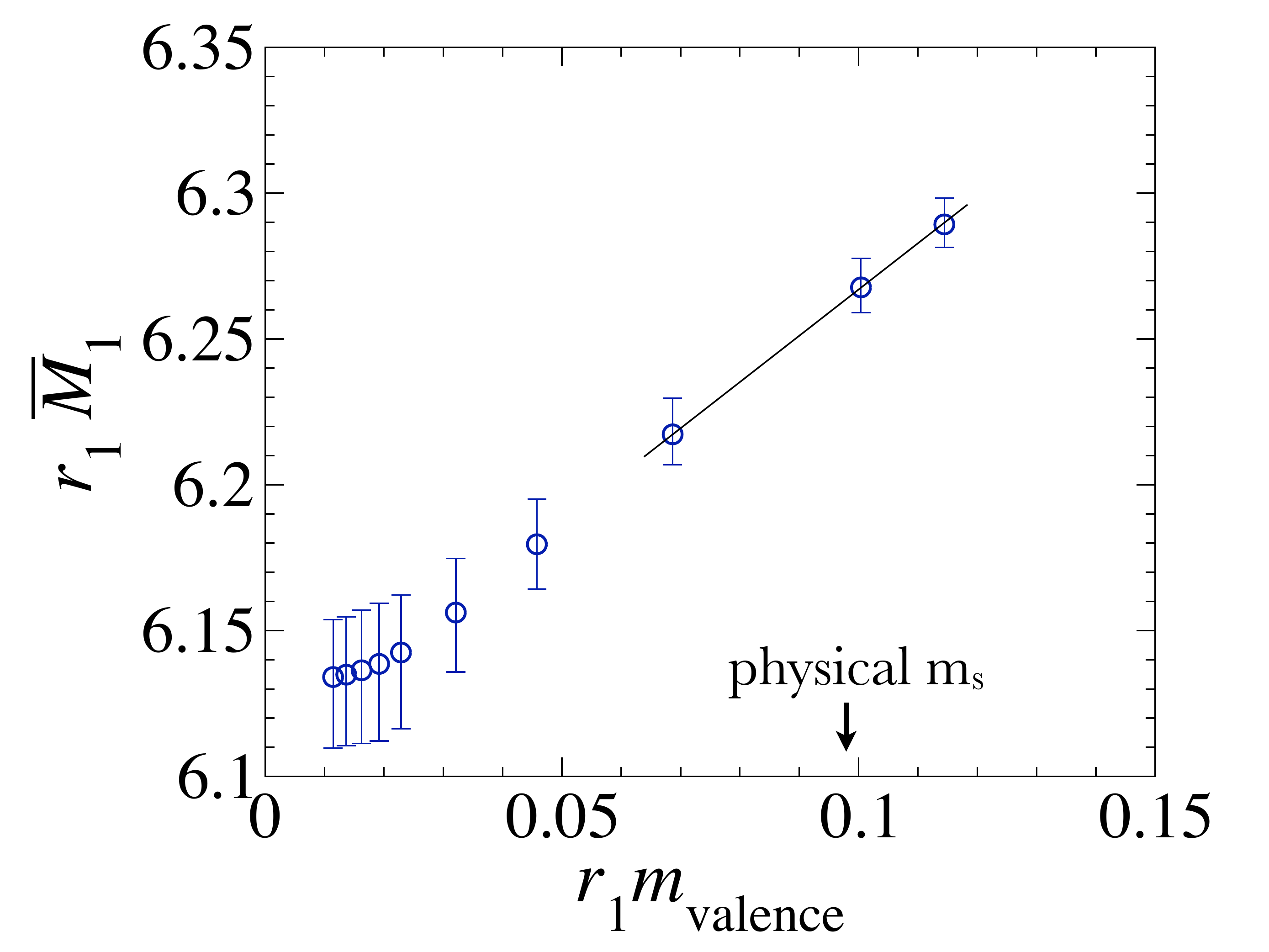} 
		\caption{Spin-averaged meson rest mass versus the light valence mass in units of $r_1$~\cite{Bernard:2000gd} on the 0.09 fm lattice.  The straight line on top of the three heaviest data points demonstrates clearly the linearity of the data near the physical strange quark mass.}
		\protect\label{valence}
	\end{center}
\end{figure}

%ensembles and mesons
We use the MILC 2+1 flavor lattices~\cite{MILC}, which have asqtad sea quarks~\cite{stag_fermion} and improved gluons~\cite{Luscher:1984xn}.  
The calculation includes three lattice spacings of approximately 0.09, 0.12, and 0.15 fm.
The ratios of the nominal up-down to strange quark masses range from $m_{u,d}/m_s = 0.1$ to 0.3 or 0.4 depending on the lattice spacing.
The heavy bottom and charm quarks are simulated using the Fermilab method~\cite{El-Khadra:1996mp}.  An asqtad quark is used for the meson's light valence quark.

%meson masses
% note we use spin-average masses
Meson rest masses are calculated using constrained curve fits to two-point correlators~\cite{Lepage:2001ym}.  We determine masses of both the pseudoscalar and vector mesons (e.g., $B_s$ and $B_s^*$) and spin average.
To determine the kinetic meson mass, $\overline{M}_2$,  we first determine the spin-averaged value of the energy at several values of momenta and then fit to the dispersion relation, Eq.~(\ref{dispreln}) for mesons.

%physical strange quark mass
Heavy-light mesons used in this calculation have a strange valence quark. This allows us to avoid a chiral extrapolation in the valence mass.  
Still, simulations are not done exactly at the physical strange mass. 
To reach it, we linearly interpolate between two neighboring points, which is validated by Fig.~\ref{valence}.

% PERTURBATION THEORY
\section {Perturbation Theory} \label{pt}
 
%Lattice PT
We use one-loop perturbation theory results to obtain the quark pole masses $m_1$ and $m_2$ from the bare mass~\cite{Mertens:1997wx, automatedPT}. 
%
%strong coupling schemes and scales
We use the $V$-scheme~\cite{LM, BLM} for the strong coupling $\alpha(q^*) = g^2(q^*)/4\pi$ and determine its value in the manner described by Mason~et~al.~\cite{Mason:2005zx}.  The scale $q^*$ should be chosen to be the typical momentum of a gluon in the loop.  To determine that momentum, we use the 
method introduced by Brodsky, Lepage and Mackenzie (BLM)~\cite{LM,BLM}  
extended to include cases where the one-loop contribution is anomalously small~\cite{HLM}.
BLM defines $q^*$ by
\begin{equation}
\label{BLM}
 \ln q^{*2} = \frac {\int d^4q \; f(q) \ln(q^2)} {\int d^4q \;  f(q)} 
\end{equation}
where $f(q)$ is the one-loop integrand.
In cases where the one-loop integral is anomalously small, or zero, this expression is inappropriate.  If the one-loop integral is zero, Reference~\cite{HLM} defines $q^*$
\begin{equation}
\label{HLM}
  \ln q^{*2} = \frac {\int d^4q \; f(q) \ln^2(q^2)} {2 \int d^4q \;  f(q) \ln(q^2)} .
\end{equation}
Solutions for small but non-zero one-loop integrals allow for a continuous transition from  Eq.~(\ref{HLM}) to Eq.~(\ref{BLM}).

%mass definitions and scales
It is known that the pole mass is not a good choice of scheme due to renormalon ambiguities~\cite{renormalons}.  
Therefore, we use a short-distance mass which is  designed to run sensibly at low renormalization scales~\cite{ElKhadra:2002wp}.  We use the potential subtracted mass~\cite{Beneke:1998rk} 
because it is additive and therefore works well with the rest-mass method as shown below.
It is based on the static quark potential and introduces a separation scale $\mu_f$. At the one-loop order
\begin{equation}
\label{PSmass}
 m_{\rm PS}(\mu_f) = m_{\rm pole} - \frac{C_F \; \mu_f}  {\pi} \frac{g^2(q^*)}{4 \pi} + O( g^4),
\end{equation}
where  $C_F = 4/3$ and
$\Lambda_{\rm QCD} < \mu_f < m_{\rm quark}$.

% PS mass
Generically, the expansion for the quark rest mass from the lattice can be written as
\begin{equation}
\label{quarkmassexpansion}
 a m_1 = a m_1^{[0]} + g^2 a m_1^{[1]} + g^4 a m_1^{[2]} +\ldots
\end{equation}
and similarly for $m_2$.
For the rest mass we can define
\begin{equation}
\label{oneloop_eqn}
 a m_{\rm 1, PS}(\mu_f) = am_1^{[0]} +  g^2(q^*) \left\{ a m_1^{[1]} - \frac{a \mu_f \, C_F}{4\pi^2}   \right\},
\end{equation}
such that Eq.~(\ref{bindingE}) yields
\begin{equation}
\label{rest_PSmass}
 	m_{\rm PS}(\mu_f) = m_{\rm 1, PS}(\mu_f) + (M_{\rm expt} - M_1),
\end{equation}
where $m_{\rm PS}(\mu_f)$ is the continuum quark mass in the potential subtracted scheme.
For $m_2$
\begin{equation}
\label{kinetic_PSmass}
 	 m_{\rm PS}(\mu_f) = am_2 \; \frac {\overline{M}_{\rm expt} }{ a \overline{M}_2}  - \frac{C_F \; \mu_f}  {\pi} \frac{g^2(q^*)}{4 \pi}.
\end{equation}

% 2-loop exploration methods
Below, our final quark masses for bottom are quoted at the scale $\mu_f$~=~2 GeV.
We reach this scale in two ways, which treat higher order effects differently.
The first uses a fixed $\mu_f$  and the calculation of  $q^*$ and  $g^2(q^*)$, as described above.
We call this the $q^*$~method.
The second has  $\mu_f'$ chosen such that the one-loop correction  is zero, e.g. for the rest-mass method $\mu_f' = 4 \pi^2 m_1^{[1]}/C_F$.  The resulting $m_{\rm PS}(\mu_f')$ is then run to the final $\mu_{f}$ by using the two-loop renormalization group equation for the PS mass:
\begin{equation}
\label{RGE}
	 	m_{PS}(\mu_f) - m_{PS}(\mu'_f) 
		= \frac{C_F}{\pi}  
			\left[
				\mu'_f \, \alpha(\mu'_f) \left\{ 1 + 2 \beta_0 \alpha(\mu'_f)    \right\}
				- \mu_f \, \alpha(\mu_f) \left\{ 1 + 2 \beta_0 \alpha(\mu_f)    \right\}
			\right] ,
\end{equation}
where $\beta_0 = \frac{1}{4\pi} (11 - 2/3 \; n_f)$.  
We call this the zero-and-run method.

 % Preliminary Results
\section{Preliminary Results} \label{results}

%new 2007
Preliminary results for the bottom quark mass were presented last year~\cite{Freeland:2006nd}.  
Several improvements have been made since then.
First, we now include the 0.09~fm lattice and omit the 0.18~fm lattice.
Second, we have added the kinetic-mass method. 
Third, we use improved scale setting for $\alpha(q^*)$~\cite{HLM},
and finally we use the zero-and-run method for the mass scale.

\begin{figure}
\begin{tabular}{cc}
	\includegraphics[scale=0.26]{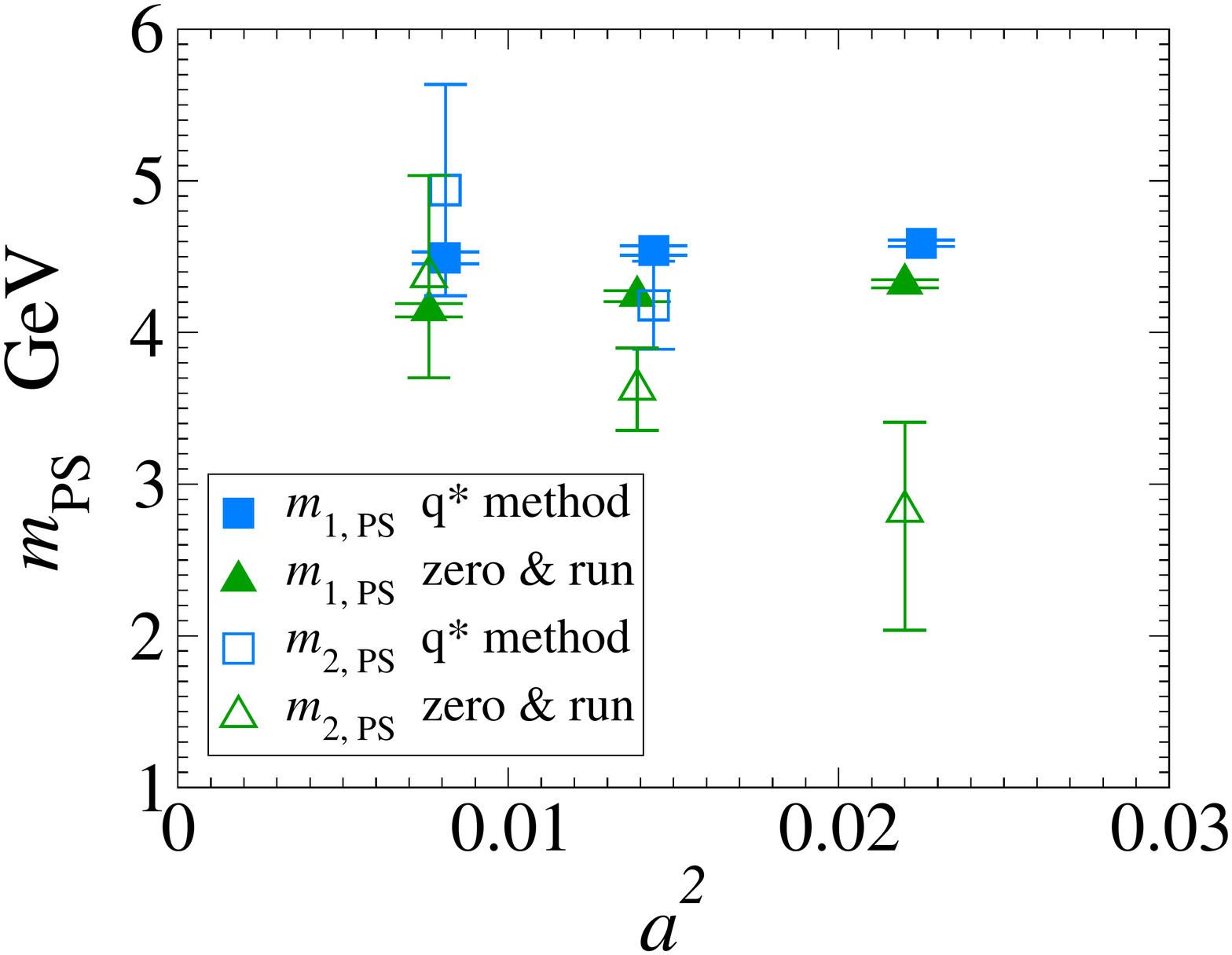} &
        \includegraphics[scale=0.26]{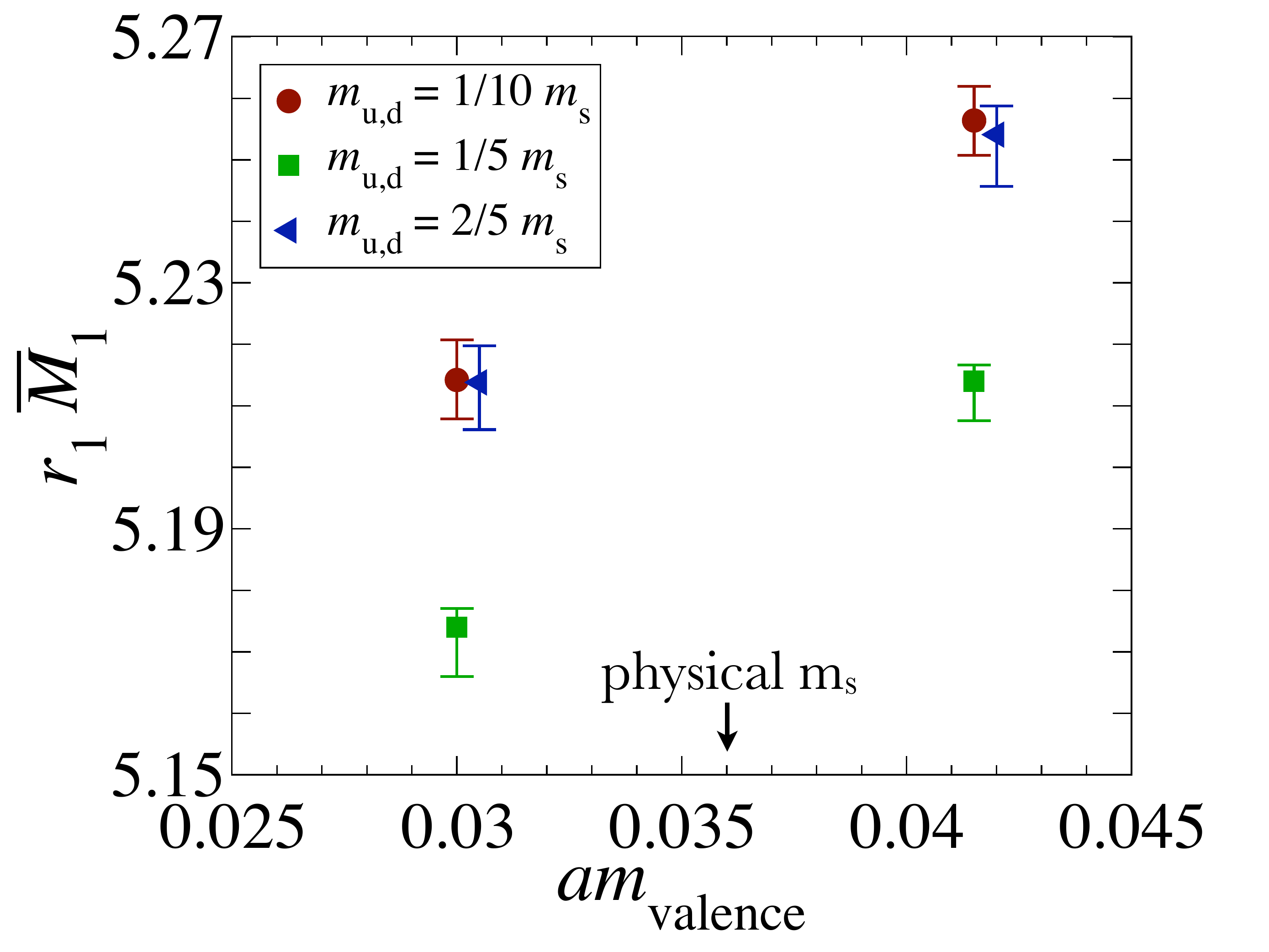}     \\
      $\quad$(a)  & $\quad$ (b)
\end{tabular}
\caption{(a)  Results for the bottom quark mass in the potential subtracted scheme for three lattice spacings.  Offsets on the $x$-axis are for clarity.  (b) Spin-averaged (rest) mass at two values of the valence mass and three different values of sea mass ratios $m_{\rm u,d} / m_{\rm s}$. Data are from the 0.12 fm lattice. Offset on the $x$-axis is for clarity}
\protect\label{mb_sea}
\end{figure}

%bottom plots
Figure~\ref{mb_sea}(a) shows updated results for the bottom quark mass.  Error bars include uncertainties from statistics, chiral sea-quark effects and the determination of the lattice spacing. 
Most noticeable are the large error bars on the kinetic mass results (open symbols) versus the rest-mass results (filled symbols).  These are primarily statistical; we hope to reduce them in the future so that the kinetic-mass method is a stronger cross-check of the rest-mass method.
%

%error budget & PT truncation error
Table~\ref{uncertainties} lists the percent uncertainties in the calculation of the bottom quark mass from the rest-mass method.  The uncertainty due to the truncation of the QCD perturbation theory clearly dominates.   
To estimate this, we take the spread in results for the quark mass from the $q^*$ and zero-and-run method on the 0.09 fm lattice.  This yields a 4\% uncertainty which is consistent with $\alpha^2(q^*) (4\pi \, m_1^{[1]})$.

\begin{table}
	\begin{center}
     	\begin{tabular}[b]{ll}   % the [b] aligns the bottom of the tabular w/ the ``text" (= bot of pdf fig)
		\hline \hline
		{\bf Source}    &       {\bf percent error} \\ \hline
		statistical                    & 0.1 \\ 
		lattice spacing determination    & 0.4 \\
		heavy-quark tuning        & 0.5 \\ 
		sea quark effects    & 0.7 \\ 
		strange mass tuning         & 0.2 \\ 
%		perturbation theory truncation    & 2 to 3 \\ 
		  perturbation theory truncation    & 4 \\
		light quarks and glue & 1 \\
		heavy quark discretization    & 0.6 \\  \hline 
%		{\bf total }             &    {\bf 3 to 4}  \\
		   {\bf total }             &    {\bf 4.3}  \\
		 %{\bf total }             &    {\bf 2.6 to 3.5}  \\
		\hline \hline \vspace{0.5ex} \\
	\end{tabular}   \\
	\caption{Percent uncertainties in the bottom quark mass.  These are added in quadrature to arrive at the total.}
	\protect\label{uncertainties}
	\end{center}
\end{table}
%

%sea quark effects
We expect the dependence on the mass of up-down sea quarks to be mild for heavy-strange meson masses.
Fig.~\ref{mb_sea}~(b) shows a typical set of meson (rest) masses for three different sea-quark ensembles.  The up-down quark mass gets smaller from 
%blue to green to red (triangle to square to circle).  
the blue triangle to the green square to the red circle.
For our central value, we use the meson mass from the ensemble with the smallest up-down quark mass and we take the largest spread in values as an estimate of the uncertainty due to sea-quark effects.
For example,  in the plot shown we use the red circle for our central value and the difference between it and the green square as an estimate of the effect of $m_{\rm u,d} \not \approx  0 $.

% discretization error
For the heavy-quark discretization error we consider only the contribution from the Darwin term,   since spin-averaging removes the hyperfine and spin-orbit interactions.
The coefficient for this term from both the 
Sheikoleslami-Wohlert 
%Clover
and continuum actions is known.  
Using the difference, $f_E$, one can estimate the uncertainty as $a^2 \, f_E \, \Lambda_{\rm QCD}^3$
 given the lattice spacing $a$ and an estimate for $\Lambda_{\rm QCD}$~\cite{Kronfeld:2003sd}.  
To get an estimate for $\Lambda_{\rm QCD}$, 
we fit two subsets of the rest-mass results (0.09 and 0.12~fm; 0.09 and 0.15~fm) to an $O(a^2)$ ansatz.  
The average of these results gives $\Lambda_{\rm QCD} = 1.3$~GeV.  
 Evaluating $a^2 \, f_E \, \Lambda_{\rm QCD}^3$ at the 0.09~fm spacing then yields a 0.6 percent error.  Although $\Lambda_{\rm QCD} = 1.3$~GeV is high, the resulting error is small and so we conservatively  take this as the uncertainty due to heavy-quark discretization.
 %The range of discretization errors quoted in Table~\ref{uncertainties} is from $a^2 \, f_E \, \Lambda_{\rm QCD}^3$ at the 0.09~fm spacing with $\Lambda_{\rm QCD}$  = 0.7 to 1.3 GeV.

% other errors  r_1, m_s, kappa
The lattice spacing (or $r_1$~\cite{Bernard:2000gd}) determination and strange mass tuning are done by the MILC Collaboration~\cite{Claude}.  We use $r_1 = 0.318(7)$~fm.  
We also include errors due to an estimated 10\%  uncertainty in the tuning of $m_s$  and an 8\% mistuning of the heavy quark.  %Numbers in the table are uncertainties from these three parameters propagated to the quark mass.
We estimate the uncertainty due to discretization of the light quarks and gluons as $\alpha \, a^2 \Lambda_{\rm QCD}^3$ and $a^4 \Lambda_{\rm QCD}^5$ with $\Lambda_{\rm QCD} = 1$~GeV and quote the larger of the two.

%  Central value
From Fig.~\ref{mb_sea}~(a), the rest-mass method clearly has much smaller errors.  In addition, the lattice spacing dependence is see to be mild.  For these reasons, we take the average of the two results from the $q*$ and zero-and-run methods for the rest-mass on the 0.09~fm lattice as our central value for the bottom quark mass.
Uncertainties in Table~\ref{uncertainties} are added in quadrature to arrive at the total.
%Using the average of the two fine lattice results from the rest-mass method as our central value, 
We have then for the bottom quark mass in the potential subtracted scheme 
%$m_{\rm b,PS}(2 \; {\rm GeV})$ = 4.46(18) GeV. 
$m_{\rm b,PS}(2 \; {\rm GeV})$ = 4.32(19) GeV.
For comparison, a QCD sum rule calculation~\cite{Pineda:2006gx} obtains  $m_{b,PS}(2 \; {\rm GeV}) = 4.52(6)$~GeV and $\overline{m}_b(\overline{m}_b) =  4.19(6)$~GeV in the $\overline{\rm MS}$ scheme.

%% Summary
%\section{Summary} \label{summary}
%We are using non-perturbative calculations of heavy-light mesons in combination with perturbative renormalization to determine the heavy-quark masses for bottom and charm.  Our results span three lattice spacing of approximately 0.09, 0.12 and 0.15 fm.  Ratios of the up-down to strange quark mass in the sea range from $m_{\rm u,d} / m_{\rm s} = 0.1$ to 0.4.  Several methods are used to estimate discretization errors and higher order perturbative corrections.  For the bottom quark mass in the potential subtracted scheme, we have a preliminary value of  $m_{\rm b,PS}(2 \; {\rm GeV})$ = 4.46(18) GeV.  The calculation for charm and conversions to the $\overline{MS}$ scheme are in progress.

% Acknowledgments
\section{Acknowledgements}
E.D.F. is supported by the M. Hildred Blewett Scholarship of the American Physical Society, www.aps.org.  
She would like to thank Fermilab for providing arrangements which allowed her to work there during her scholarship year,
 and M. Hildred Blewett for her generosity in funding such a scholarship.
USQCD computer resources were used for these calculations.
%Fermilab is operated by Fermi Research Alliance, LLC, under Contract No.~DE-AC02-07CH11359 with the United States Department of Energy.

\end{document}